\documentclass[twocolumn]{article}
\usepackage{amssymb}

\usepackage{graphicx}
\usepackage{amsmath}

\input{tcilatex}

\begin{document}

\title{NON-PERTURBATIVE \thinspace QCD\thinspace\ SPIN STUDIES\thanks{%
Plenary talk presented by T.P. Cheng at SPIN-98, Protvino, Russia, Sept.
8-12, 1998.}}
\author{T. P. Cheng$^{\dagger }$ and Ling-Fong Li$^{\ddagger }$ \\
%EndAName
$^{\dagger }${\small Department of Physics and Astronomy, University of
Missouri, St. Louis, MO 63121 USA}\\
$^{\ddagger }${\small Department of Physics, Carnegie Mellon University,
Pittsburgh, PA 15213 USA}}
\date{}
\maketitle

\begin{abstract}
A whole class of non-perturbative QCD studies (\emph{e.g.} the instanton
models, chiral quark models, \emph{etc}.) indicates that the effective
degrees of freedom for the physics in the low $Q^{2}\lesssim 1\,GeV^{2}$
region could be the constituent quarks(CQs) and internal Goldstone bosons
(IGBs). This leads to a nucleon structure with the spin being carried by
three constituent quark systems, each composed of a massive compact CQ
surrounded by a $q\bar{q}$ sea perturbatively generated by the valence
quark's IGB emissions. Such a CQ-system has a total angular momentum of 1/2
and a small anomalous magnetic moment, built up from a quark-spin
polarization and a significant orbital motion in the quark sea. The
distinctive phenomenological signal for such a non-perturbative structure is
that the polarization of the sea-quarks differs from that of the antiquarks: 
$\left( \Delta _{q}\right) _{\text{sea}}\neq \Delta _{\bar{q}}\,:$ the
sea-quarks are polarized negatively, $\left( \Delta _{q}\right) _{\text{sea}%
}<0,$ while the antiquarks are not polarized, $\Delta _{\bar{q}}=0.$ This
picture also suggests a negligibly small gluon polarization, $\Delta G\simeq
0$. All such features can be tested by experiments in the near future.
\end{abstract}

\section{Introduction}

The QCD gauge coupling is not small for the distance and energy scales
involved in the structural study of hadrons composed of light $u,\,d,\,$and $%
s$ quarks. Thus questions such as ``what carries the proton spin?'' will
necessarily involve an understanding of non-perturbative QCD. Even if we
were able to answer this spin question in term of percentages of the proton
spin being carried by its component quarks and gluons, to explicate the
mechanism that built up the final spin from these fundamental QCD degrees of
freedom (DOF) is still a daunting task. For this we may ultimately rely on
such approaches as lattice QCD calculations. But instead of such a head-on
attack (and in order to have a simple physical understanding), it may well
be useful to adopt a two-stage approach using an effective DOF description
of the non-perturbative phenomena.

\subsection{The two-stage approach}

At the first stage, one tries to find the effective degrees of freedom for
the strong interactions in this low $Q^{2}$ regime, in terms of which the
physics description will be simple, intuitive, and phenomenologically
correct. At the second stage, one then looks for the relation of these
non-perturbative effective DOF and the QCD quarks and gluons.

An example of the non-perturbative DOF is the constituent quark of the
non-relativistic quark model. The constituent quarks $U,\,D,\,$and $S$ carry
the same quantum numbers as the QCD Lagrangian quarks $u,\,d,\,$and $s$, but
they have much greater mass values. It is well-known that the
non-relativistic quark model correctly describes most of the static hadronic
properties. The spectroscopy and baryon magnetic moments are well fitted%
\footnote{%
For a recent pedigogical review, see \cite{CL-sch97}.} with CQ masses: 
\begin{equation*}
M_{U}\simeq M_{D}\equiv M_{U,D}\simeq 350\,MeV,\;M_{S}\simeq 500\,MeV
\end{equation*}
which are to be compared to the Lagrangian quark mass values (at $1\,GeV$)
of 
\begin{equation*}
m_{u,d}\equiv \frac{m_{u}+m_{d}}{2}\simeq 6\,MeV,\;\;\;m_{s}\simeq 150\,MeV.
\end{equation*}
Presumably the extra mass of the CQ results from some non-perturbative QCD
interaction. Thus, at the second stage, one needs to work out the details
showing how QCD can endow the light quarks with a sizable mass --- yet
retaining their simple Dirac magnetic moment structure, $\mu
_{Q}=q_{Q}/2M_{Q}.\;$Later on we shall discuss examples of\ models (such as
the chiral quark model and the instanton model) that can account for such
entities.

The advantage of separating the non-perturbative study into two stages is
that the effective DOF of the first stage provide us with a simple common
language, which can facilitate, and stimulate, communications among
different groups investigating the nucleon structure problem, and which
allows simpler ways to generalize to more difficult non-perturbative
phenomena. Furthermore, alternative approaches have many features in common
at the first stage. (What sets them apart are some of the details and
precise connection to the underlying QCD.) Thus this two stage approach
should also help us in making comparisons of different theories.

\subsection{The naive quark sea}

The non-relativistic quark model is often referred to as being the naive
quark model (NQM). Relativistic corrections have not been taken into
account, and it lacks a $q\bar{q}$ quark sea, which is expected in any
quantum field theoretical description involving quarks.

The NQM gives a good description of the baryon magnetic moment (\emph{%
i.e.\thinspace }minimally affected by the quark sea). This led people to
believe that the NQM account of the quark contribution to the baryon spin
should also be reliable\footnote{%
Quark contribution $\Delta q$ to the proton spin is the sum of quark and
antiquark polarizations, $\Delta q=\Delta _{q}+\Delta _{\bar{q}},$ while the
magnetic moment involves their difference $\widetilde{\Delta q}=\Delta
_{q}-\Delta _{\bar{q}}$ \ --- because antiquark has the opposite electric
charge. In NQM there is no antiquark in the proton, hence $\widetilde{\Delta
q}=\Delta q$ and the magnetic moment result directly probes the quark spin
contributions $\Delta q.$}. Thus the discrepancy between the NQM prediction
for the neutron axial-vector constant $g_{A}=\Delta u-\Delta d=5/3$ and the
experimental value of $1.26\simeq 5/4$ has generally been attributed to a
possibly large relativistic correction\cite{Rel-corr, Rel-Brodsky}. But such
a reduction depends on the details of the quark momentum distribution inside
the nucleon.

Clearly one would identify the NQM quarks as the valence quarks. Therefore,
the recent study of the hadron structure can be said to be primarily the
study of the quark sea. But, the conventional expectation of the quark sea
properties had been heavily influenced by perturbative QCD reasoning:
Namely, the sea is pictured to result from the quark-pair production by
gluons.

\begin{itemize}
\item  The sea is supposed to have the same amount of $\bar{u}$ and $\bar{d}$%
, because gluon is flavor-independent, and $u$ and $d$ quarks have similar
masses.

\item  The gluon coupling is such that quarks and antiquarks are expected to
have the same polarization, (to the extend that masses can be ignored): 
\begin{equation*}
\left( \Delta _{q}\right) _{\text{sea}}=\Delta _{\bar{q}}\;\;\text{or\ \ }%
\left( \Delta _{q}\right) _{\text{sea}}-\Delta _{\bar{q}}\equiv \left( 
\widetilde{\Delta q}\right) _{\text{sea}}\simeq 0.
\end{equation*}
Since $\widetilde{\Delta q}$ enters into the magnetic moment calculation,
this possibility was thought to be helpful to account for the absence of a
quark sea effect on the baryon moments.

\item  One part of the conventional expectation that is non-perturbative in
character is the suggestion that OZI rule\cite{OZI} should be applicable in
general. Namely, not only we expect a suppression of $s\bar{s}$ in the
vector channel (which explains, for example, the ideal mixing observed among
the vector meson spectrum) but in other channels as well, \emph{e.g.} in the
scalar and axial vector channels: thus the suppression of the proton matrix
elements of 
\begin{equation*}
\left\langle p\left| \bar{s}s\right| p\right\rangle \simeq \left\langle
p\left| \bar{s}\gamma _{\mu }\gamma _{5}s\right| p\right\rangle \simeq 0.
\end{equation*}
This has led to the anticipation of a small pion-nucleon sigma term\cite
{Cheng76}, and $\Delta s\simeq 0$\cite{EJsum}$.$
\end{itemize}

All such conventional expectations have been called into question by
experimental measurements.

\section{Experimental measurements}

\subsection{The proton spin structure}

The interest in the proton spin problem has been high, ever since the
discovery in the late 1980's by the European Muon Collaboration\cite{EMC}
that the Ellis-Jaffe sum rule\cite{EJsum} is violated by the experimental
data from polarized deep inelastic scattering. It implies that the proton
quark sea has a polarized strange quark component, $\Delta s\neq 0.$ This is
followed by more polarized DIS experiments, SMC at CERN\cite{SMC}, E142-3,
E154-5 at SLAC\cite{SLAC-spin}, and HERME at DESY\cite{HERMES}, which
generally support the original EMC findings. Here we quote a typical set of
phenomenological result as reported by SMC\cite{SMC-quote}: 
\begin{eqnarray}
\left( \Delta u\right) _{\text{expt}} &=&0.82\pm 0.02,\;\;\;\left( \Delta
d\right) _{\text{expt}}=-0.43\pm 0.02,  \notag \\
\left( \Delta s\right) _{\text{expt}} &=&-0.10\pm 0.02,\;\;\;\left( \Delta
\Sigma \right) _{\text{expt}}=0.29\pm 0.06,  \notag \\
&&\;\;  \label{deltq-expt}
\end{eqnarray}
all evaluated at $Q^{2}=5\,GeV^{2}.$ These results have been obtained with
the assumption of flavor-SU(3) symmetry\cite{EJsum}. There are indications%
\footnote{%
This is indicated both by models for SU(3) breaking\cite{su3br-axial} and by
more direct measurement by SMC\cite{SMC-semi}.} that SU(3) breaking
corrections might lower somewhat the magnitude of $\Delta s$. Here we note
the main feature that they deviate significantly from the NQM prediction: 
\begin{eqnarray}
\left( \Delta u\right) _{\text{NQM}} &=&\frac{4}{3},\;\;\;\left( \Delta
d\right) _{\text{NQM}}=-\frac{1}{3},  \notag \\
\left( \Delta s\right) _{\text{NQM}} &=&0,\;\;\;\;\;\;\left( \Delta \Sigma
\right) _{\text{NQM}}=1.  \label{deltq-NQM}
\end{eqnarray}
This comparison naturally leads one to the possible interpretation\footnote{%
This interpretation implicitly assumes that the relativistic correction is
not the principal reduction factor of the NQM values to the ones in (\ref
{deltq-expt}). For example, this approach would attribute the reduction $%
\left( 5/3\rightarrow 5/4\right) $\ of $g_{A}$ mainly to the depolarization
effect of the quark sea.} of Eqs.(\ref{deltq-expt}) and (\ref{deltq-NQM}) as
indicating a negatively polarized quark sea. Namely, if we identify the NQM
values as the polarization of the valence quarks, and the difference as due
to the $q\bar{q}$ quark sea: 
\begin{equation}
\left( \Delta q\right) _{\text{expt}}=\left( \Delta q\right) _{\text{NQM}%
}+\left( \Delta q\right) _{\text{sea}}\;\;\text{with\ \ }\left( \Delta
q\right) _{\text{sea}}<0.  \label{val-sea-sum}
\end{equation}
The quark sea is polarized in the direction opposite to the proton spin.

\subsection{The proton flavor structure}

The spin puzzle is part of nucleon structure problem and should not be
treated in isolation. There is by now a considerable amount of experimental
data having bearing on the flavor structure of the proton.

\begin{itemize}
\item  The are more $\bar{d}$\ than $\bar{u}$\ in the proton: NMC\cite{NMC}
first discovered that the Gottfried sum rule\cite{GSR} of $l$-$N$ DIS was
violated. This can be translated into the statement about the difference of
antiquark density as $\bar{d}-\bar{u}=0.147\pm 0.026.\;$That there are more $%
\bar{d}$ than $\bar{u}$ has also been confirmed by comparing data from $pp$
with $pn$ Drell-Yan processes, first by NA51 at CERN\cite{NA51} and more
recently by Fermilab E866\cite{E866}.

\item  The\ presence of $s\bar{s}$ in the nucleon is not suppressed by the
basic mechanism that generates the quark sea, even though it is reduced
somewhat by the strange quark mass factor $M_{S}>M_{U,D}.\;$This point is
illustrated by the contrasting phenomenological results of the pion-nucleon
sigma term $\sigma _{\pi N}$\ (which is the pion-nucleon scattering
amplitude at a particular kinematic point\cite{CD71}) and neutrino charm
production\cite{CCFR}. An SU(3) analysis\cite{Cheng76} of the $\sigma _{\pi
N}$\cite{sigma91} (entirely similar to the Ellis-Jaffe calculation\cite
{EJsum} of the flavor spin factors) suggests a surprisingly large fraction
of strange quarks (strange quark and antiquark number divided by the sum of
all quark numbers) in the nucleon $f_{s}\simeq 0.18,\;$which is reduced to
about $0.10$ by SU(3) breaking effects\cite{sig-su3br}. Neutrino charm
production is also sensitive to the presence of strange quarks in the
nucleon. However its phenomenological fits suggest \cite{str-fun-fits} a
ratio of $2\bar{s}/\left( \bar{u}+\bar{d}\right) \simeq 0.5.$
\end{itemize}

\section{The chiral quark idea}

Non-perturbative QCD has two prominent features: confinement and dynamic
breaking of chiral symmetry. The key to understand both is the structure of
QCD vacuum. The most detailed theoretical model, (\emph{i.e.} covering most
of the non-perturbative issues in a self-consistent way), has been the
instanton approach of Diakonov, et al.\cite{Diak}, or the related work by
Shuryak, et al.\cite{Schur} At a more phenomenological level, there are also
the various quark models based on Nambu-Jona-Lasinio interaction\cite
{Hatsuda-pr}. All these models share the feature of having a ``chiral
quark'' effective theory based on the interaction among constituent quarks
and some entities having the quantum numbers of Goldstone bosons\footnote{%
In the instanton models, there are actually no free propagating $0^{-}$
states at all, they are just short-hands for $q\bar{q}$ loop effects: $\bar{q%
}q$ pairs ``propagate'' by leaping among states associated with instantons.}%
. Since the basic chiral quark idea is simpler to explain in the form as
first formulated by Manohar and Georgi\cite{MG-cqm}, this is the language we
shall adopt in this presentation. Later on we shall comment on the possible
problems of this version of the chiral quark model $\left( \chi QM\right) $
and its common features with the instanton approach.

The chiral quark idea is based on the possibility that chiral symmetry
breaking $\left( \chi SB\right) $ takes place at a distance scale much
smaller than the confinement radius: in terms of the energy scales, $\Lambda
_{\chi \text{SB}}\gg \Lambda _{\text{conf}}\approx \Lambda _{\text{QCD}}\,$\
with $\Lambda _{\chi \text{SB}}\simeq 1\,GeV$ vs $\Lambda _{\text{QCD}%
}\simeq 0.1-0.2\,GeV.\;$As distance increases, the increased QCD coupling
strength $g$ will be such that the non-perturbative phenomenon of $\chi SB$
is triggered around $\Lambda _{\chi \text{SB}}^{-1}$, way before the hadron
exterior $\Lambda _{\text{conf}}^{-1}.$ In this interior region, the QCD
vacuum acquires a non-trivial structure with a $q\bar{q}$ condensate: $%
\left\langle \bar{q}\left( x\right) q\left( x\right) \right\rangle _{0}\neq
0,\;$and pseudoscalar massless states, the Goldstone bosons, come into being%
\footnote{%
For a textbook review of the strong interaction chiral symmetry, see, for
example, \cite{CLbook}.}. Thus in the hadron interior, but not so small a
distance that perturbative QCD is applicable, the effective degrees of
freedom are the CQs and internal GBs.

\begin{itemize}
\item  \underline{Constituent quarks}: They are just the ordinary QCD
quarks, but now propagating in the non-trivial QCD vacuum having $q\bar{q}$
condensate. The quarks pick up an extra mass through the interaction with
the condensate: 
\begin{equation}
M_{\text{Q}}=m_{\text{q}}+\frac{f}{\Lambda _{\chi \text{SB}}^{2}}%
\left\langle \bar{q}q\right\rangle _{0}+...
\end{equation}
If for example $\left. M_{\text{Q}}\right| _{m_{\text{q}}=0}=f\left\langle 
\bar{q}q\right\rangle _{0}/\Lambda _{\chi \text{SB}}^{2}+..\simeq 350\,MeV$
we can reproduce CQ masses $M_{\text{U,D}}$ and $M_{\text{S}}$ in the ranges
of $350$ and $500\,MeV$, respectively. This mechanism of mass generation is
very similar to that of the electroweak symmetry breaking, where the
non-trivial vacuum corresponds to the presence the Higgs condensate and the
generation of the (Lagrangian) masses for leptons and quarks. The fermions
gaining masses through such a mechanism are compact objects having a size on
the order of symmetry breaking scale. That the constituent quarks are
compact (because of a small $\Lambda _{\chi \text{SB}}^{-1}\ll \Lambda _{%
\text{conf}}^{-1}$) is supported by the absence of any observed excited
quark states\cite{Weinberg90}. This also implies that a CQ does not have an
anomalous moment.

\item  \underline{Internal Goldstone bosons}: One of the direct consequence
of spontaneous chiral symmetry breaking is that there will be massless
pseudoscalar states. When propagating outside the confinement radius, they
are the familiar light pseudoscalar octet mesons, $\pi ,\;K,$ and $\eta .$
Here we are discussing Goldstone bosons in the hadron interior. To emphasize
that they may well have different propagation properties, \emph{e.g.}
different effective masses (also see footnote 5), we call them the internal
Goldstone bosons (IGB)
\end{itemize}

What about the gluons? Of course, it is the QCD gluonic interaction that
brings about the non-perturbative phenomena of chiral symmetry breaking and
the generation of massless pseudoscalar bound $q\bar{q}$ Goldstone bosons.
In the $Q^{2}\lesssim 1\,GeV^{2}$ range, besides such ``non-perturbative
gluons'', no perturbative gluons are expected to be an\ important factor.
[For further comment on gluonic contribution, see Sec.5.3] Namely, after $%
\chi SB$ has taken place, the interaction should be dominated by that among
the CQs and IGBs. This remanent interaction is expected to be much weaker
than the original QCD quark gluon interaction. Schematically, we can think
of this in terms of the Hamiltonian being\ \thinspace $\mathcal{H}_{\text{QCD%
}}=\mathcal{H}_{\text{massless q,g}}+V_{int}\;$at the short distances, while
at distance scale longer than $\Lambda _{\chi \text{SB}}^{-1},\;$one has the
effective Hamiltonian\thinspace $\;\mathcal{H}_{\text{effective}}=\mathcal{H}%
_{\text{massless GB, massive Q}}+v_{int}^{\prime }\;$with $v_{int}^{\prime
}\ll V_{int}.\;$Otherwise the non-perturbative interaction among the quarks
and IGB would be such as to completely obscure these particle identities.
Because the quarks are so heavy in this regime, we expect the probability
for processes producing the sea $q\bar{q}$ pairs to be small, making the
reaction effectively perturbative. [This is also \emph{a posteriori}
justified in $\chi QM$ calculations.]

In short, while the fundamental QCD interaction, in terms of quarks and
gluons, is non-perturbative in the low energies, after separating out the
non-perturbative effects of $\chi SB,$ the remanent interaction (in terms of
IGB and massive CQs) is again perturbative. In this way one can consider the
following simple (tree diagram) mechanism for the quark sea generation
through the perturbative emission of IGB by a valence quark $Q$: 
\begin{equation}
Q_{\uparrow }\longrightarrow GB+Q_{\downarrow }^{\prime }\longrightarrow Q+%
\bar{Q}^{\prime }+Q_{\downarrow }^{\prime }  \label{sea-cqm-mech}
\end{equation}
The subscript arrow indicates the helicity of the quark; no arrow means an
unpolarized quark.

Whatever created by the quantum fluctuation, \emph{e.g. }all quarks other
than the original valence quarks in the NQM, is considered to be part of the
sea. Thus all three final-state quarks in (\ref{sea-cqm-mech}) are taken to
be components of the quark sea.

\subsection{Spin structure in the $\protect\chi QM$}

The axial $\gamma _{5}$ coupling of IGB to the constituent quark reduces, in
the non-relativistic limit, to $\left( \vec{s}/M\right) \cdot \vec{p}$ where 
$\vec{s}$ and $\vec{p}$ are the spin and momentum operators, respectively.
The one-pseudoscalar-meson-exchange between CQs has the same spin dependent $%
\left( \vec{s}_{i}/M_{i}\right) \cdot \left( \vec{s}_{j}/M_{j}\right) $
structure as that due to the one-gluon-exchange\cite{DGG-glue} --- except
that the overall color factor is replaced by some flavor coefficients. Thus
all the successful account of the spin dependent features in the baryon
spectrum by the gluonic exchange mechanism (\emph{e.g.} the mass differences
between $\Delta $ and $N,$ between $\Sigma $ and $\Lambda $ etc.) can be
taken over by the IGB exchange description. In fact, as shown by Glozman and
Riska\cite{Glozman}, the IGB scenario leads to a better spectroscopy:
certain features on the level orderings in the higher mass strange baryon
states can now be explained. Moreover, the mystery of why there is\ not a
significant spin-orbit contribution can also be accounted for by this IGB
exchange mechanism.

This same spin-dependence of the IGB-quark coupling also implies that the GB
vertex will flip the helicity of the quark $Q_{\uparrow }\rightarrow
Q_{\downarrow }^{\prime }$ in the emission process (\ref{sea-cqm-mech}).
Furthermore, the final $Q+\bar{Q}^{\prime }$ quarks in (\ref{sea-cqm-mech}),
being produced through the GB channel must have their spins add up to a
spin-zero system,\ $2^{-1/2}\left( Q_{\uparrow }\bar{Q}_{\downarrow
}^{\prime }-Q_{\downarrow }\bar{Q}_{\uparrow }^{\prime }\right) .$

\begin{itemize}
\item  \emph{Antiquarks are not polarized}\cite{CL96}\emph{\ --- }Since the
antiquarks in the sea must all be produced through such spin-zero mesons,
the antiquarks in the quark sea are not polarized, $\Delta _{\bar{q}}=0$ ---
the probability for finding an spin-up antiquark equals to that for an
antiquark in the spin-down state.

\item  \emph{Quark sea is negatively polarized --- }Since both $Q$ and $\bar{%
Q}^{\prime }$ quarks are unpolarized, the polarization of the entire final
state (the quark sea) must be given by the $Q_{\downarrow }^{\prime }$
quark, which is opposite to the initial quark helicity state. This naturally
leads to a negatively polarized sea, in qualitative agreement with
phenomenological observation, as expressed in Eq.(\ref{val-sea-sum}) so that 
\begin{equation}
\Delta \Sigma =\Delta \Sigma _{\text{val}}+\Delta \Sigma _{\text{sea}}
\label{del-sig-sum}
\end{equation}
is less than one.

\item  \emph{Quark sea has a positive orbital angular momentum}\cite
{CL98-prl}\emph{\ --- }Angular momentum conservation in the reaction $%
Q_{\uparrow }\rightarrow GB+Q_{\downarrow }^{\prime }$ requires that $\left[
GB,\;Q_{\downarrow }^{\prime }\right] $ be in the relative P-wave state, $%
\left\langle l_{z}\right\rangle =+1,$ in order to compensate the $%
Q_{\uparrow }\rightarrow Q_{\downarrow }^{\prime }$ quark helicity flip $%
\delta \sigma =-2,$%
\begin{equation}
%TCIMACRO{\UNICODE[m]{0xbd}}%
%BeginExpansion
{\frac12}%
%EndExpansion
\Delta \Sigma _{\text{sea}}+\left\langle L_{z}\right\rangle =0
\label{ang-mom-consv}
\end{equation}

\item  \emph{The proton spin is built up from quark spins and orbital motion
in the quark sea --- }Combining Eqs.(\ref{del-sig-sum}), (\ref{ang-mom-consv}%
) and $\Delta \Sigma _{\text{val}}=1,$\emph{\ }we see that the proton spin
sum $\left( \equiv 1\right) $ is composed of quark spin and orbital angular
momentum terms: 
\begin{equation}
\Delta \Sigma +2\left\langle L_{z}\right\rangle =1.
\end{equation}
Many authors have suggested the possibility of proton spin receiving a
significant contribution from the orbital angular momentum\cite{orbit-others}%
. However, most of such schemes have $\left\langle L_{z}\right\rangle $
arising from the valence quarks in higher orbital states (so called
configuration mixing). In contrast, the present discussion concerns the
orbital angular momentum in the quark sea. To make this distinction clearer,
it may be helpful to consider the valence CQ and the sea it generates
together as a ``\emph{constituent quark system}''. What we suggest here is
that, even though the sea has a depolarization effect, it is compensated by
its positive orbital motion --- so that the whole CQ-system remains to be a
spin 
%TCIMACRO{\UNICODE{0xbd}}%
%BeginExpansion
$\frac12$%
%EndExpansion
\ entity. Proton's three CQ-systems themselves, however, remain in the
relative S-wave state, \emph{i.e.} not in any significant relative orbital
motion.
\end{itemize}

\subsection{Magnetic moments in the $\protect\chi QM$}

The quark sea will also contribute to the baryon magnetic moment\ $\mu _{%
\text{B}}=\mu _{\text{val}}^{\text{B}}+\left( \mu _{\text{sea}}^{\text{B}%
}\right) .\;$Because $\left( \Delta _{q}\right) _{\text{sea}}-\Delta _{\bar{q%
}}=\left( \widetilde{\Delta q}\right) _{\text{sea}}\neq 0,$ it is puzzling
why the NQM can yield a good account of the baryon magnetic moments even
when the sea is strongly polarized. The answer is that, because the spin
polarization and orbital angular momentum have opposite signs, Eq.(\ref
{ang-mom-consv}), their magnetic moment contributions also tend to cancel 
\cite{CL98-prl}: 
\begin{equation}
\left( \mu _{\text{sea}}^{\text{B}}\right) =\mu _{\text{spin}}^{\text{B}%
}+\mu _{\text{orb}}^{\text{B}}\simeq 0.  \label{mu-cancel}
\end{equation}
This is indicated in a non-relativistic calculation. While the
non-relativistic approximation is useful in giving us a simple intuitive
physical picture, the relativistic field-theoretical loop calculation will
automatically include both the spin and orbital contributions. It yields an
anomalous magnetic moment for the constituent quark system. The claimed
cancellation in (\ref{mu-cancel}) simply means that the anomalous moment due
to the chiral field is particularly small. This is indeed the case as shown
in the explicit calculations by Dicus et al.\cite{Dicus} and by Brekke\cite
{Brekke}.

Thus we see that even though the quark sea has a strong depolarization
effect, the constituent quark system still has spin 
%TCIMACRO{\UNICODE{0xbd}}%
%BeginExpansion
$\frac12$%
%EndExpansion
\ with an approximate Dirac moment. This explains the reason why NQM can
yield a good account of the baryon magnetic moments even though its spin
content predictions has been found to be incomplete.

The anomalous moment is small, but measurable. It has been shown\cite{Brekke}
that the NQM fit of the baryon moments can be considerably improved by
giving quarks small anomalous moments and the fit supplemented by exchange
current contributions, which arise naturally in the $\chi QM.$

\subsection{Flavor structure in the $\protect\chi QM$}

Here we make the brief comment that the chiral quark approach can naturally
account for the observed proton flavor puzzles: $\bar{d}\gg \bar{u}$ and a
significant $\bar{s}.$

\begin{itemize}
\item  In the $\chi QM$ the valence $u$ quark, through the intermediate
state of IGB, is more likely to produce $\bar{d}$, and valence $d$ tends to
produce $\bar{u}.$ While $u\rightarrow \pi ^{+}d\rightarrow u\bar{d}d$ \
and\ $u\rightarrow K^{+}s\rightarrow u\bar{s}s$\ are allowed, the $%
u\rightarrow \pi ^{-}q^{5/3}$ or $K^{-}q^{5/3}\rightarrow \bar{u}....$
processes cannot take place because there does not exit a $q^{5/3}$ quark
having a $5/3$ charge. This naturally leads a proton quark sea having more $%
\bar{d}$ than $\bar{u}$ because there are two valence $u$ quarks and only
one valence $d$. (This effect is diluted somewhat by $q\rightarrow \pi
^{0}q,\,\eta ^{0}q$ which produce equal amount of $\bar{d}$ and $\bar{u}.)$

\item  While there is a tendency for $u\nrightarrow \bar{u}...$ and $%
d\nrightarrow \bar{d}...\,$both valence $u$ and $d$ can produce $\bar{s}.$
If one could ignore the effect of $M_{S}>M_{U,D}$, there would be more\
strange than non-strange antiquarks: $\bar{s}>$ $\bar{d}$, $\bar{u}$ in the
quark sea\cite{CL95}. The $M_{S}$ suppression effect is not expected to be
overwhelming because constituent mass difference is considerably smaller
than current quark mass differences, $\left( M_{S}/M_{U,D}\right) \ll \left(
m_{s}/m_{u,d}\right) ,$ and because of non-vanishing internal momentum $%
\left\langle k^{2}\right\rangle \neq 0.$
\end{itemize}

\section{Chiral QM calculations}

Bjorken\cite{BJ-cqm} was the first to suggest that the chiral quark idea may
be relevant to a solution of the proton flavor and spin puzzles. Eichten,
Hinchliffe and Quigg\cite{EHQ} carried out the non-relativistic quark model
calculation (as well as the chiral field calculations of the $x$ and $Q^{2}$
distributions). Cheng and Li\cite{CL95} have proposed to work with a $\chi
QM $ including not only the octet GBs but also the singlet $\eta ^{\prime }$
meson, with a broken $U(3)\rightarrow SU(3)\times U\left( 1\right) $
symmetry. In the leading $N_{c}^{-1}$\ expansion ($N_{c}$ is the number of
colors), there are nine Goldstone bosons, but the concomitant U(3) symmetry
must be broken by higher order $N_{c}^{-1}$ corrections (so as to account
for the presence of axial anomaly). This is implemented by allowing for
distinctive couplings $f_{1}\left( \neq f_{8}\right) $ for the singlet GB$.$

The phenomenology at the SU(3) symmetric level suggests a negative ratio $%
f_{1}/f_{8}$\cite{CL95}$.$ Since the fit is not very sensitive to precise
value, we shall simply fix it at $f_{1}=-f_{8}.$ In this way, there left
only one parameter to adjust. This is the transition probability $a\propto
\left| f_{8}\right| ^{2}$ for the emission process $u\rightarrow \pi ^{+}d$
and its SU(3) equivalents. The model calculation results are shown in the
third columns of Tables 1 and 2. At the next stage we can include the broken
SU(3) effects of $M_{S}>M_{U,D},$ or equivalently $m_{K,\eta }>m_{\pi },$ by
introducing suppression factors for strange-quark containing GB amplitudes 
\cite{Song-su3br, CL98-prd}. To see whether our approach has the
qualitatively correct phenomenological features, we will limit the number of
extra parameters to one by applying a common factor of $\left[ \left\langle
k^{2}\right\rangle +m_{GB}^{2}\right] ^{-1}$ for each GB amplitude. The
results from the two parameter, $a$ and $\,\left\langle k^{2}\right\rangle ,$
calculation\cite{CL98-prd} are shown in the last columns of Tables 1 and 2.
We also recall that the same model can give a good account of the magnetic
moment data as well as the baryon spectroscopy.

These calculations have not included relativistic corrections \cite
{Rel-Brodsky}. Relativistic reduction of the spin fractions is likely to
improve the agreement, but to include this one would need more parameters
representing the internal momentum distributions.

At this stage the calculated densities are understood to be averaged over
all $x.$ [See further discussion in the next Section.] Since this is
non-perturbative model calculation, the results are for $Q^{2}\lesssim
1\,GeV^{2}.\;$But the comparison with the experimental data measured at $%
Q^{2}\simeq 5\,GeV^{2}$ is still valid because experimentally it is known
that the $Q^{2}$\ variation is small in going from 1 to 5 $GeV^{2}.$ Putting
it another way, if we had been calculating the distribution functions, they
would be the ``initial distributions'' at $Q^{2}\approx 1\,GeV^{2},$ from
which the higher $Q^{2}$ distributions can be deduced by performing the pQCD
calculations.

\begin{center}
\begin{tabular}{|c|c|rl|}
\hline
SPIN & pheno. values &  & $\chi QM$ \\ \hline
&  & \multicolumn{1}{|c}{SU$_{3}$} & \multicolumn{1}{|c|}{br'n SU$_{3}$} \\ 
\cline{3-4}
$\Delta u$ & $\;0.82\pm 0.02\;\;\;\;\;$ & \multicolumn{1}{|c}{$\;0.78$} & 
\multicolumn{1}{|c|}{$\;0.85$} \\ 
$\Delta d$ & $-0.43\pm 0.02\;\;\;\;\;$ & \multicolumn{1}{|c}{$-0.33$} & 
\multicolumn{1}{|c|}{$-0.40$} \\ 
$\Delta s$ & $-0.10\pm 0.02\,\left( \downarrow ?\right) $ & 
\multicolumn{1}{|c}{$-0.11$} & \multicolumn{1}{|c|}{$-0.07$} \\ 
$\Delta \Sigma $ & $\;0.29\pm 0.06\;\;\;\;\;$ & \multicolumn{1}{|c}{$\;0.34$}
& \multicolumn{1}{|c|}{$\;0.38$} \\ 
$\Delta _{\bar{u}},\;\Delta _{\bar{d}}$ & $\;0.01\pm 0.07\;\;\;\;\;$ & 
\multicolumn{1}{|c}{$0$} & \multicolumn{1}{|c|}{$0$} \\ 
$g_{A}$ & $1.257\pm 0.03\;\;\;$ & \multicolumn{1}{|c}{$1.12$} & 
\multicolumn{1}{|c|}{$1.25$} \\ 
$F/D$ & $0.575\pm 0.016\;\;\;$ & \multicolumn{1}{|c}{$2/3$} & 
\multicolumn{1}{|c|}{$0.57$} \\ \hline
\end{tabular}
\end{center}

\begin{quotation}
\underline{Table 1}.{\small \ Comparison of }$\chi QM${\small \ spin
structure results with phenomenological values. The third column corresponds
to the SU(3) symmetric result with a single parameter }$a=0.12,${\small \
while the fourth column for the broken SU(3) result with two parameters: }$%
a=0.15${\small \ and }$\left\langle k^{2}\right\rangle =350\,MeV^{2}.\;$%
{\small The quark contributions to the proton spin }$\left( \Delta q,\Delta
_{\bar{q}}\right) ${\small \thinspace are from an SU(3) symmetric analysis
by SMC\cite{SMC-quote, SMC-semi}. The }$\Delta s${\small \ magnitude may be
reduced when broken SU(3) effect is taken into account (see footnote 3).
Hence the symbol }$\left( \downarrow ?\right) ${\small \ behind the stated
value. }$F/D$ is the ratio of SU(3) reduced matrix elements for the axial
currents.
\end{quotation}

\begin{center}
\begin{tabular}{|c|c|rl|}
\hline
FLAVOR & pheno.\ values &  & $\chi QM$ \\ \hline
& \ \ \ \ \ \ \ \ \ \ \ \ \ \ \ \ \ \ \ \ \ \  & \multicolumn{1}{|c}{SU$_{3}$%
} & \multicolumn{1}{|c|}{br'n SU$_{3}$} \\ \cline{3-4}
$\bar{d}-\bar{u}$ & $0.147\pm 0.026$ & \multicolumn{1}{|c}{$0.15$} & 
\multicolumn{1}{|c|}{$0.15$} \\ 
$\sigma _{_{\pi N}}:f_{s}$ & $0.18\overset{m_{\text{{\tiny s}}}}{\rightarrow 
}0.10$ & \multicolumn{1}{|c}{$0.19$} & \multicolumn{1}{|c|}{$0.09$} \\ 
$2\bar{s}/\left( \bar{u}+\bar{d}\right) $ & $\simeq 0.5$ & 
\multicolumn{1}{|c}{$1.86$} & \multicolumn{1}{|c|}{$0.6$} \\ \hline
\end{tabular}
\end{center}

\begin{quotation}
\underline{Table 2}. {\small Comparison of }$\chi QM${\small \ flavor
structure results with phenomenological values. Exactly the same parameters
have been used as in Table 1. The }$\bar{u}-\bar{d}${\small \ difference is
from NMC measurement which is supported by more recent E866 data, although
the central value is somewhat lower. The strange quark fraction }$f_{s}$%
{\small \ is deduced from pion-nucleon sigma term with both SU(3) symmetric
and breaking values given, while the antiquark ratio in the last row is from
structure function fits based primarily on the CCFR data on neutrino charm
production}.
\end{quotation}

\section{Discussion}

These simple calculations seem to bolster the idea that CQ and internal GB
as effective DOF can yield an adequate description of the principal features
of the proton's spin/flavor structure. This provides the supporting reason
as why the simple constituent quark model works. The massive but compact
constituent valence quarks are surrounded by a quark sea, which is
perturbatively generated by IGB emissions.

The advantage of working with such effective DOF description is that it is
simple enough so that generalization to more complicated and difficult
physical situations will be possible. Chiral quark model has already been
applied by Troshin and Tyurin to hadron scattering, in particular in their
attempt to calculate the single spin asymmetries in $pp$ and $p\bar{p}$
collisions\cite{T-T}.

As we have already mentioned, this chiral effective description is shared by
a number of non-perturbative approaches: instanton\cite{Diak, Schur, Koch},
quark model with NJL interaction\cite{Hatsuda-pr}, chiral bag model\cite
{chi-bag} and skyrmion model\cite{skyrme}. Some of them, because further
approximation, (\emph{e.g.} skyrmion model of baryon being the large-$N_{c}$
approximation of QCD), may lead to a phenomenology somewhat different from
that discussed above. For example, the skyrmion model has the prediction of $%
\Delta \Sigma =0$\cite{BEK}.

\subsection{The instanton approach}

As we have emphasized, the main feature in the chiral quark approach is the
dynamical chiral symmetry breaking, leading to a non-trivial QCD vacuum
inside the hadron. Not surprisingly the most complete model is the one that
has the most worked-out mechanism of chiral symmetry breaking. This is the
instanton approach.

Instantons are classical Euclidean solutions of QCD\cite{Belavin}.
Physically one may think of them as patches of intense gluon field
fluctuations in spacetime (because they are $\sim g^{-1}$) that arise from
tunnelling among different classical QCD vacua\cite{tHooft}. Attempts to
understand the hadron structure by assuming an instanton dominance in
non-pQCD phenomena were made at the earliest stage of instanton research\cite
{CDG}. The recent study has been greatly stimulated by the discovery, first
through phenomenological study of correlations\cite{Schur}, and then by
variational principle calculation\cite{Diak}, that large instanton effects
are suppressed. The infrared problem is now under control.

Starting from the only dimensional parameter in QCD, namely $\Lambda _{\text{%
QCD}}$, the two basic characteristic length scales of the instanton vacuum
can be obtained through variational calculation: the average distance
between neighboring instantons $\bar{R}\simeq 1\,fm,$ and their average mean
square radius $\bar{\rho}\simeq 0.35\,fm.\;$This in turn allows one to
deduce quantities such as $\left. M_{Q}\right| _{m_{q}=0}\simeq 350\,MeV$
and $\left\langle \bar{q}q\right\rangle _{0}\simeq -\left( 250\,MeV\right)
^{3},\;etc.$ We remark that these two scales $\left( \bar{R},\bar{\rho}%
\right) $ essentially replaced the two scales of$\;\left( \Lambda _{\text{%
conf}},\Lambda _{\chi \text{SB}}\right) $ in the simple $\chi QM.$

The study of the equation of motion for light quarks propagating in the
instanton field shows the existence of fermionic ``zero modes'', \emph{i.e.}
quarks localized around instanton. A well defined procedure\cite
{Banks-Casher} then leads one to conclude that there are quark pair
condensate in the ground state: $\left\langle \bar{q}q\right\rangle _{0}\neq
0,$ hence, a spontaneously broken chiral symmetry. Furthermore, 't Hooft has
shown that (for the three light flavor case) an effective six-quark
interaction, in the determinantal form, is induced, 
\begin{equation}
\mathcal{H}_{G}=\det_{i,j}\left[ \bar{q}_{iR}q_{jL}+h.c.\right]
\label{det-int}
\end{equation}
where the flavor indices $i,j=1,2,3.$ Because this term is symmetric under $%
SU_{L}(3)\times SU_{R}(3)$ but not under $U_{A}\left( 1\right) \;$it will
give a mass to the singlet would-be-Goldstone boson, thus solving the axial $%
U_{A}\left( 1\right) $ problem\cite{tHooft}.

Instanton-induced quark-interaction is qualitatively similar to that of the
chiral quark. Forte, Dorokhov, and Kochelev\cite{Forte, Koch}\ have pointed
out that just as the GB-quark coupling, the instanton induced quark
interaction flips the quark chirality. The 't Hooft interaction (\ref
{det-int}) implies that an instanton absorbs a left-handed quark of each
flavor and emits a right-handed quark of each flavor, $\bar{u}_{R}u_{L}\bar{d%
}_{R}d_{L}\bar{s}_{R}s_{L}.\;$This provides a mechanism for produce a
negatively polarized quark sea, and (in the equal mass limit) a flavor
structure of $\bar{s}>\bar{d}>\bar{u}$ in the proton.

\subsection{Alternative approaches}

If different theories yield similar effective DOF descriptions, how do we
differentiate them apart? We can do so by the following considerations:

\emph{Theoretical consideration: }We give two\ illustrative examples: (i)
The meson cloud model\cite{cloud-models}, at one level, is very similar to
the chiral quark model: in both cases the pseudoscalar mesons play a central
role. In the cloud model, the IGBs of $\chi QM$ are replaced by the physical
pseudoscalar mesons exterior to the nucleon. The DIS processes proceed
through the Sullivan mechanism, \emph{i.e.} the lepton probe scatters off
the meson cloud surrounding the target nucleon\cite{Sullivan}. The flavor
asymmetry is thought to result from the excess of $\pi ^{+}$ (hence $\bar{d}$%
) compared to $\pi ^{-},$ because $p\rightarrow \pi ^{+}+n,$ but $%
\nrightarrow $ $\pi ^{-}+...$ if the final states are restricted to the
nucleons. But, it still has to be worked out why the long-distance feature
of the pion cloud surrounding the nucleon should have such a pronounced
effect on the DIS which should probe the interior of the target. Also, such
meson cloud effect may well be reduced by emissions such as $p\rightarrow
\pi ^{-}+\Delta ^{++},$ etc. (ii) The $\chi QM$ as formulated by Manohar and
Georgi\cite{MG-cqm} appears to have two sets of pseudoscalars: The GBs
appearing in the effective Lagrangian as independent DOF, as well as another
set of pseudoscalar bound $\bar{Q}Q$ states. On the other hands, the chiral
effective Lagrangian from the instanton approach does not have an explicit
kinetic energy term for GBs: there is no propagating GB in the instanton
effective theory (see footnote 5).

\emph{Consideration of model details: }Even though different theories can
result in similar effective structure at the quark model level, they may
very well differ when more detailed phenomenology is examined. (i) An
obvious example is that while several theories may give the same densities,
when averaged over all $x$ as discussed in Sec.4. Different theories have
different distributions. Thus the distribution from chiral effective theory
having propagating GBs will be different from theories such as instanton
theory, where IGBs are just short-hand for $\bar{q}q$ loop effects\footnote{%
The instanton theory can apparently produce quark distributions in general
agreement\cite{Diak-sf} with those extracted from phenomenology\cite{Reya}.}%
. (ii) While the singlet and octet coupling ratio $f_{0}/f_{8}$ in the $\chi
QM$ calculations is an arbitrary parameter. The instanton determinantal
interaction suggests that it should be negative\cite{CL-Koch} (an effect
directly related to the singlet meson gaining a mass even in the chiral
limit). This is in accord with the phenomenology as shown in Sec.4.

Clearly much work need to be done in filling out the details of the model
predictions which should then be checked by experimental measurement.

\subsection{The gluonic contribution}

The proton spin problem is often discussed in the context of percentage of
the spin being carried by quark spin $\Delta \Sigma $, gluon spin $\Delta G$%
\cite{gluon-spin} and the orbital angular momentum contributions: 
\begin{equation}
%TCIMACRO{\UNICODE[m]{0xbd}}%
%BeginExpansion
{\frac12}%
%EndExpansion
\Delta \Sigma +\Delta G+\left\langle L_{z}\right\rangle =%
%TCIMACRO{\UNICODE[m]{0xbd}}%
%BeginExpansion
{\frac12}%
%EndExpansion
.
\end{equation}
How is our $\chi QM$ conclusion, valid for $Q^{2}\lesssim 1\,GeV^{2},$\
related to such an expression at the higher $Q^{2}$ region? As we have
emphasized, the constituent quarks are just the ordinary quarks propagating
in the range of $Q^{2}\lesssim 1\,GeV^{2}.\;$In the same manner, one would
expect the non-perturbative gluons of the low $Q^{2}$ range to be the
starting point for $\Delta G$ \ to evolve to higher $Q^{2}$ as prescribed by
pQCD. The $\chi QM$ suggests: 
\begin{equation}
%TCIMACRO{\UNICODE[m]{0xbd}}%
%BeginExpansion
{\frac12}%
%EndExpansion
\Delta \Sigma +\left\langle L_{z}\right\rangle =%
%TCIMACRO{\UNICODE[m]{0xbd}}%
%BeginExpansion
{\frac12}%
%EndExpansion
,\;\;\;\Delta G=0.
\end{equation}
About one-third of proton spin resides in the quark constituents while about
two-thirds in the orbital angular momentum of the quark sea. There is no $%
\Delta G$ in the low energy regime. Namely, the non-perturbative gluons
which bring about $\chi SB$ and confinement are not polarized. Thus one
would expect a negligibly small $\Delta G$ even for the higher $Q^{2}$
region. This is consistent with the experimental observation of the a very
weak $Q^{2}$-dependence in $\Delta \Sigma $, which is determined by the $%
\Delta G\left( Q^{2}\right) $ contribution via the axial anomaly equation.

\subsection{Concluding remarks}

The nucleon spin/flavor structure, from the pQCD perspective, has been
thought as being puzzling. This naturally suggests that the non-pQCD in the
low energy $Q^{2}\lesssim 1\,GeV^{2}$ region as being the source for this
non-trivial structure. In this presentation we have concentrated on the
possibility of using the effective DOF of constituent quarks and internal
Goldstone bosons for a simple description of the non-perturbative phenomena.
This gives the initial distributions which are to be evolved to the higher $%
Q^{2}$ ranges through pQCD equations.

The chiral quark model is able to provide us with an unified account for the
main features of the observed nucleon spin and flavor structure. The simple
picture is that nucleon is composed of three CQ systems which are the
valence quarks being surrounded by their quark sea. The sea is generated
perturbatively through the IGB emissions. The $u$ valence system has a
significant $\bar{d}$ and $\bar{s}$ components and the $d$ valence system
has a significant $\bar{u}$ and $\bar{s}$, with the would-be-large strange
component being counterbalanced by mass suppression factor. The quark sea
has a strong negative spin polarization which is compensated by a
significant orbital motion so that each CQ system has spin 1/2 and a small
anomalous magnetic moment. While its property of 
\begin{equation}
\left( \Delta \Sigma \right) _{\text{sea}}=-2\left\langle L_{z}\right\rangle
\simeq -2/3
\end{equation}
is compatible with the known spin and magnetic data, its distinctive
phenomenological prediction of no antiquark polarization,$\,$ and a
negligibly small gluonic polarization, 
\begin{equation}
\Delta _{\bar{q}}=0,\;\;\;\text{and}\;\;\;\Delta G\simeq 0
\end{equation}
can be checked by future experimental measurements, in particular the
various spin programs at HERA\cite{HERA-pol}, RHIC\cite{RHIC-spin}, and
COMPASS\cite{COMPASS}.\medskip

\textbf{Acknowledgment: }\emph{One of us (T.P.C.) wishes to thank members of
the UC-Santa Cruz Institute for Particle Physics for the warm hospitality
shown him during the 1998 summer, when this report was prepared. L.F.L.
acknowledges the support from U.S. Department of Energy (Grant No.
DOE-Er/40682/127).}

\end{document}